\documentclass{article}
\usepackage[utf8]{inputenc}
\usepackage{times}
\usepackage{amsfonts}
\usepackage{amsmath,amssymb}
\usepackage{indentfirst}
\usepackage{wrapfig}
\usepackage{graphicx}
\usepackage{graphics,color}
\usepackage{caption}
\usepackage[toc,page]{appendix}
\usepackage{color}
\usepackage{cite}
\usepackage[inactive]{srcltx}
\usepackage[T1]{fontenc}
\usepackage{float}
\usepackage{hyperref}
\usepackage{subfigure}

\begin{document}
	\date{}
\begin{center}
	{\Large\bf Generation and transfer of entangled states between two connected microtoroidal cavities: 
analysis of different types of coupling}
\end{center}
\begin{center}
	{\normalsize Emilio H. S. Sousa$^{(a,b)}$, A. Vidiella-Barranco$^{(b)}$\footnote{vidiella@ifi.unicamp.br} and J. A. Roversi$^{(b)}$}
\end{center}
\begin{center}
	{\normalsize{$^{a}$``Professora Alda Façanha'' State School of Professional Education}}\\
	{\normalsize{ 61700-000, Aquiraz, CE, Brazil}}\\
	{\normalsize{$^{b}$Gleb Wataghin Institute of Physics - University of Campinas}}\\
	{\normalsize{ 13083-859   Campinas,  SP,  Brazil}}\\
\end{center}
\begin{abstract}
We investigate the generation and transfer of entangled states between two coupled microtoroidal 
cavities considering two different types of couplings, namely i) via a bridge qubit and ii) via evanescent 
fields. The cavities support two counter-propagating whispering-gallery modes (WGMs) 
that may also interact with each other. We firstly show that it is possible to transfer, with high fidelity, 
a maximally entangled state between the two modes of the first cavity (cavity $1$) to the two modes of 
the second cavity (cavity $2$), independently of the type of coupling. Interesting differences, though, 
arise concerning the generation of entangled states from initial product states; if the cavities are coupled 
via a bridge qubit, we show that it is possible to generate a 4-partite entangled state involving all four cavity 
modes. On the other hand, contrarily to what happens in the qubit coupling case, it is possible to generate bipartite 
maximally entangled states between modes of different cavities from initial separable states for cavities coupled by 
evanescent waves. Besides, we show that different entangled states between the propagating and counter-propagating modes 
of distinct cavities may be generated by tuning the interaction between modes belonging to the same cavity (intra-cavity 
couplings). Again, this is possible only for the couplings via evanescent waves. For the completion of our work, we 
discuss the effects of losses on the dynamics of the system.
\end{abstract}
%
\section{Introduction}
Quantum entanglement is a fundamental resource and also indispensable in many important applications such as 
quantum key distribution \cite{RevModPhys.74.145,diamanti2016practical}, quantum computing \cite{arute2019quantum}, 
dense coding \cite{PhysRevLett.93.210501}, and quantum information processing \cite{stajic2013future}. For instance,
the performance of tasks such as quantum teleportation in a quantum network, depends on both the control of 
entanglement generation and the transfer of entangled states between distant parties \cite{wu2017fast}.
Several theoretical and experimental works have shown that cavity QED systems promote an efficient route towards the generation and 
distribution of entanglement \cite{sousa2019selective,Blythe_2006,kato2019observation,kimble2008quantum}.
In \cite{PhysRevA.70.022320}, the authors have demonstrated a process of entanglement transfer engineering, where two remote qubits 
interact with an entangled two-mode continuous-variable field. Whereas the authors in \cite{PhysRevA.90.062342} used a single 
superconducting qubit to induce entanglement between two separated transmission line resonators, in \cite{PhysRevA.86.012318} the 
authors introduce two gap-tunable superconducting qubits to generate a two-mode entangled state through dissipation. 
The generation of two-mode entangled states using superconducting qubits has also been studied in
\cite{PhysRevLett.105.050501,PhysRevA.85.022335}, while in \cite{kim16} the authors proposed the use of a bridge 
qubit to couple cavities containing each one a qubit. We remark that the coupling via a bridge qubit would allow the 
adjustment of the interaction between the two microtoroidal cavities (including to switch it on/off),
which is of importance for the scalability of quantum circuits \cite{kim16}. 
In \cite{PhysRevA.64.050301}, the authors used a single circular Rydberg atom to prepare two-modes of the field of a superconducting 
cavity in a maximally entangled state, while in \cite{wang2011two} the authors have proposed the creation of maximally entangled states 
for two modes via a single atom resonating with an ultra-high $Q$ microtoroidal cavity. The preparation of a steady entangled state 
between two separated nitrogen-vacancy (NV) centers attached to the outer surface of two microtoroidal cavities coupled by a 
WGM field has been proposed in \cite{PhysRevA.100.052332}. In \cite{PhysRevA.85.042306}, the authors showed that it 
is possible to produce a steady entangled state of the two NV centers with two WGM microresonators, coupled 
either by an optical fiber-taper or via evanescent fields. 

To our knowledge though, the investigation and comparison of different couplings between two WGMs microtoroidal cavities aiming at the
entanglement generation and transfer of two-mode entangled states has not yet been addressed in the literature. Here, we report protocols of 
generation and transfer of entangled states between two coupled microtoroidal cavities, considering setups with two different 
types of inter-cavity couplings, namely: i) a bridge qubit; and ii) evanescent waves. Each one of the cavities supports two 
counter-propagating WGMs, clockwise and counterclockwise propagating modes, that may be coupled to each other due the cavity 
imperfections \cite{sousa2019role,PhysRevA.83.023803}. We are interested in the investigation of the effects 
of different couplings on the dynamics of entanglement of the WGMs in the cavities, as well as 
in the control of the generation of entangled states by tuning the parameters of the system. We show that it is possible 
to transfer, with high fidelity, a maximally entangled state between the two-modes of the first cavity to the modes of the second cavity for 
each one of the considered couplings. In particular, the transfer of entangled states between cavities is independent of the 
interaction between the intra-cavity modes if the cavities are coupled via evanescent waves.
Concerning the possibilities of generation of entangled states from initial product states, we find that in 
the case of the evanescent waves coupling, bipartite maximally entangled states involving different modes may 
be generated, depending on the interaction between the intra-cavity modes. This allows the selection of the generated states 
by tuning the aforementioned interactions. Yet, such flexibility is not possible if the cavities are coupled via a bridge qubit. Nonetheless, we show that precisely in the case of coupling via a qubit, it is possible to generate a 4-partite entangled state involving the four cavity field states. Thus, in this system of two microtoroidal cavities, each of the cavity-cavity couplings considered here has particular features which would allow the transfer and generation of a wide range of entangled states of the quantized field. For completeness, we investigate the influence of losses on the dynamics of entanglement in the system.

Our paper is organized as follows. Firstly, in Section \ref{sec:model}, we present the models and descriptions of the physical 
systems. In Section \ref{sec:results}, we investigate and discuss the transfer and generation of entanglement (quantified by the Concurrence) 
for different types of couplings and different initial states. The effects of losses are 
discussed in Section \ref{sec:dissip}. Finally, we summarize our conclusions in Section \ref{sec:conclu}.

\section{Models of two coupled microtoroidal cavities}
\label{sec:model}

Consider a system composed of two separate toroidal cavities, each one supporting two whispering gallery counter-propagating modes (WGM’s) of frequencies $\omega_{ci}$ 
$(i=1,2)$ with corresponding photon annihilation (creation) operators given by $a_{i}$ ($a_{i}^\dagger$) and $b_{i}$ ($b_{i}^\dagger$). 
We also assume an interaction between the two WGMs (with intra-cavity coupling constants $J_{i}$) induced by small deformations 
in the toroid \cite{sousa2019role,PhysRevA.83.023803}. Hence, the Hamiltonian of such a system will read:

\begin{eqnarray}
	H_{S} &=& \hbar \omega_{c1} (a_{1}^\dagger a_{1} + b_{1}^\dagger b_{1}) + \hbar \omega_{c2} (a_{2}^\dagger a_{2} + b_{2}^\dagger b_{2}) \nonumber \\
	&+& \hbar J_{1}(a_{1}^\dagger b_{1} + b_{1}^\dagger a_{1} ) + \hbar J_{2}(a_{2}^\dagger b_{2} + b_{2}^\dagger a_{2}).
	\label{eq:Hs}
\end{eqnarray}

Here, we suppose that the cavities can be connected using two different types of coupling (one at a time), 
namely: i) via a bridge qubit, i.e., a two-level system, which can be an artificial atom, a nitrogen 
vacancy center, etc., having transition frequency $\omega_{a}$; and ii) via evanescent waves. 

\begin{figure}[htbp]
	\centering
	\includegraphics[scale=0.6]{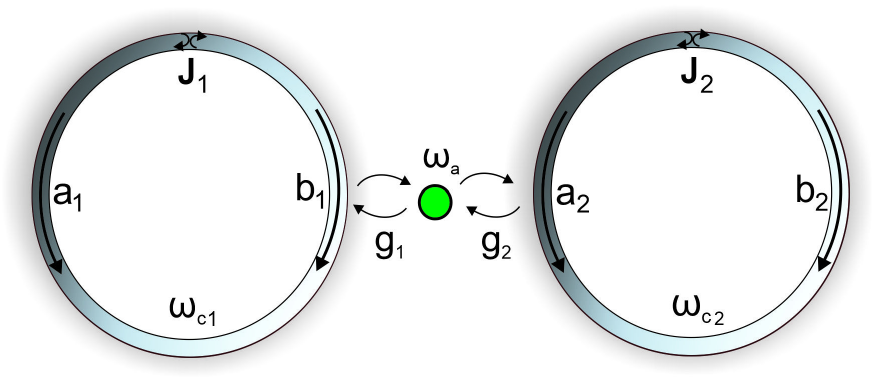}
	\caption{A schematic illustration showing the setup of two toroidal cavities supporting modes 
		$a_1 \wedge b_1$ and $a_2 \wedge b_2$, coupled via a brigde qubit, with coupling constants
		$g_1 = g_2 \equiv g$.}
	\label{fig_qubit}
\end{figure}

In the bridge qubit case, as shown in Fig. \ref{fig_qubit}, we consider that the two WGMs counter-propagating modes of each cavity are 
simultaneously coupled to the qubit with coupling constants $g_i$\footnote{We assume atoms symmetrically coupled to the two WGMs, and we make $g_{1,2} \equiv g$.}. 
Thus, the total Hamiltonian with the additional interaction term has the form:
\begin{eqnarray}
	H_{T} = H_{S} + H_{g},\label{eq:Hq}
\end{eqnarray}

where
\begin{eqnarray}
	H_{g} &=& \hbar \omega_{a}\sigma^{+} \sigma^{-} + \hbar g_{1}( a_{1}^{\dagger} \sigma^{-} + a_{1} \sigma^{+}) 
	+ \hbar g_{1}( b_{1}^{\dagger} \sigma^{-} + b_{1} \sigma^{+}) \nonumber  \\ 
	&+& \hbar g_{2}( a_{2}^{\dagger} \sigma^{-} + a_{2} \sigma^{+}) + \hbar g_{2}( b_{2}^\dagger \sigma^{-} + b_{2} \sigma^{+}). 
	\label{hamilqubitcoup}
\end{eqnarray}
Here $\sigma^{+}$ and $\sigma^{-}$ are the raising and lowering operators of the qubit. 

\begin{figure}[htpb]
	\centering
	\includegraphics[scale=0.6]{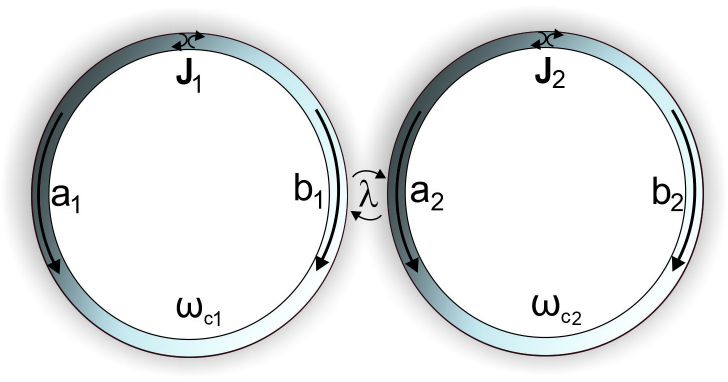}
	\caption{A schematic illustration showing the setup of two toroidal cavities supporting modes 
		$a_1 \wedge b_1$ and $a_2 \wedge b_2$, coupled directly via evanescent waves, with coupling constant $\lambda$.}\label{fig_waves}
\end{figure}

If the two cavities are coupled via evanescent fields as shown in Fig. \ref{fig_waves}, the total Hamiltonian with the additional term 
has the form:
\begin{eqnarray}
	H_{T} = H_{S} + H_{\lambda},\label{eq:Hlambda}
\end{eqnarray}

where
\begin{eqnarray}
	H_{\lambda} =  \hbar \lambda (e^{- i \phi} a_{1}^{\dagger} b_{2} + e^{ i \phi} b_{2}^{\dagger} a_{1} 
	+ e^{- i \phi} b_{1}^{\dagger} a_{2} +  e^{ i \phi} a_{2}^{\dagger} b_{1}). 
\end{eqnarray}
Here $\lambda$ is the effective coupling constant between the connected cavities. Such a direct coupling is due to a 
coherent photon exchange (at a rate $\lambda$) that may occur in several optical systems \cite{evers11}. The phase $\phi$  
is related to the propagation distance between the microtoroids. In order to avoid delaying effects in the photon propagation, 
a short distance limit between the toroids should be imposed (here we take $\phi=0$).

\section{Results: transfer and generation of entangled states}
\label{sec:results}

In what follows, we will discuss the transfer of entanglement as well as the generation of entangled states in the 
two-cavity system, considering each one of the couplings presented above. 

\subsection{Entangled state transfer}

We would like now to investigate under which conditions we are able to transfer an entangled state of the field modes in cavity $1$ 
to the field modes in cavity $2$. Firstly we consider the case of coupling via a bridge qubit with the following initial conditions: 
the two modes of cavity $1$ prepared in an entangled state, the qubit in its ground state and the two modes of cavity $2$ 
in their vacuum states\footnote{The cavity fields are denoted as
	$\vert 00\rangle_{c1} = \vert 0\rangle_{a_1}\otimes \vert 0\rangle_{b_1}$ and $\vert 00\rangle_{c2} = \vert 0\rangle_{a_2}\otimes \vert 0\rangle_{b_2}$. For simplicity, we assume $J_1 = J_2 = J$ and $g_1 =g_2 = g$. }
\begin{equation}
	\vert\Psi(0)\rangle_{g} = (\cos \theta\vert 10\rangle + \sin \theta \vert 01\rangle)_{c1}\vert g\rangle \vert 00\rangle_{c2}.\label{eq:initialstqubit}
\end{equation}

Our transmission protocol should work in such a way that entanglement would be 
transferred from cavity $1$ to cavity $2$ in the following way:
\begin{eqnarray}
	\label{eq:initial_states}
	\vert \Psi(0)\rangle_{g} &=& \left(\cos \theta\vert 10\rangle + \sin \theta \vert 01\rangle_{c1}\right)\vert g \rangle \vert 00 \rangle_{c2} \\ 
	&\Rightarrow& \vert \Psi'\rangle_{g} = \vert 00 \rangle_{c1}\vert g\rangle \left(\cos \theta \vert 10\rangle + \sin \theta \vert 01\rangle\right)_{c2}.\nonumber
\end{eqnarray}

Thus, if we depart from the initial state in Eq.($\ref{eq:initialstqubit}$), the time-evolved 
state of the system using the Hamiltonian in Eq. (\ref{eq:Hq}) (coupling via a bridge qubit) is
\begin{equation}
	|\Psi (t)\rangle_{g} = G_1 (t)|10g00\rangle + G_2 (t)|01g00\rangle + G_3(t)|00e00\rangle + G_4(t)(|00g10\rangle + |00g01\rangle),
\end{equation} 
where
\begin{eqnarray}
	G_1 (t) = \frac{1}{4}  \left[ {2\emph{e}}^{-\alpha t} (\cos\theta - \sin\theta) +  \Gamma + \beta \right],  
\end{eqnarray}
\begin{eqnarray}
	G_2 (t) = \frac{1}{4}  \left[ {2\emph{e}}^{-\alpha t} (\sin\theta - \cos\theta) +  \Gamma + \beta \right],
\end{eqnarray}
\begin{eqnarray}
	G_3 (t) = \frac{2\textit{i}\Gamma \emph{e}^{-\alpha t/2} \sinh (\Delta^2 gt/2)}{\Delta^2}    
\end{eqnarray}
and
\begin{eqnarray}
	G_4 (t) &=& \frac{(\cos\theta + \sin\theta)\emph{e}^{(\alpha - \Delta^2 g) t/2}}{8g\Delta^2} \\ \nonumber
	&\times& \left[ \textit{i}(\emph{e}^{\Delta^2 gt} -1)J + (1 + \emph{e}^{\Delta^2 gt} - 2\emph{e}^{(\alpha + \Delta^2 g)t/2} )\Delta^2 g \right],
\end{eqnarray}
with
$
\alpha = \textit{i}J, \qquad
\Gamma = \emph{e}^{\alpha t}(\cos\theta + \sin\theta), \qquad 
\Delta = \sqrt{\textit{i}[16 + (J/g)^2]} \qquad
$
and
\begin{equation}
	\beta = \frac{1}{\Delta^2} \left\lbrace  \emph{e}^{\alpha t/2} (\cos\theta + \sin\theta) \left[ \Delta^2\cosh(\Delta^2 gt) + \textit{i}J \sinh(\Delta^2 gt)/g  \right]   \right\rbrace.   
\end{equation}

For the coupling via evanescent waves, the initial state will read
\begin{equation}
	\vert \Phi(0)\rangle_{\lambda} = (\cos \theta \vert 10\rangle + \sin \theta \vert 01\rangle)_{c1}\vert 00\rangle_{c2},\label{eq:initialstothers}
\end{equation}
with the corresponding transmission protocol
\begin{eqnarray}
	\vert \Phi(0)\rangle_{\lambda} &=& \left(\cos \theta\vert 10\rangle + \sin \theta \vert 01\rangle\right)_{c1}\vert 00 \rangle_{c2} \\ \nonumber &\Rightarrow& 
	\vert \Phi'\rangle_{\lambda} = \vert 00 \rangle_{c1} \left(\cos \theta\vert 10\rangle + \sin \theta \vert 01\rangle\right)_{c2}.
	\label{eq:initial_states1}
\end{eqnarray}
In this case (evanescent waves) with initial state in Eq.(\ref{eq:initialstothers}), the time-evolved 
state of the system using the Hamiltonian in Eq. (\ref{eq:Hlambda}) reads
\begin{equation}
	|\Phi (t)\rangle_{\lambda} = W_1 (t)|1000\rangle + W_2(t)|0100\rangle + W_3(t)|0010\rangle + W_4(t)|0001\rangle,
\end{equation}
where 
\begin{equation}
	W_1(t) = \cos(\lambda t) [\cos\theta \cos(J\lambda t) + \textit{i}\cos\theta\sin(J \lambda t)],
\end{equation}

\begin{equation}
	W_2(t) = \cos(\lambda t) [\sin\theta \cos(J\lambda t) + \textit{i}\sin\theta\sin(J \lambda t)],
\end{equation}

\begin{equation}
	W_3(t) = \sin(\lambda t) [\textit{i}\cos\theta \cos(J\lambda t) - \sin\theta\sin(J \lambda t)],
\end{equation}
and
\begin{equation}
	W_4(t) = \sin(\lambda t) [\textit{i}\cos\theta \sin(J \lambda t) - \cos\theta\sin(J \lambda t)].
\end{equation}

\begin{figure}[htpb]
	\centering
	\subfigure[\label{fig:2a}]{\includegraphics[scale=0.5]{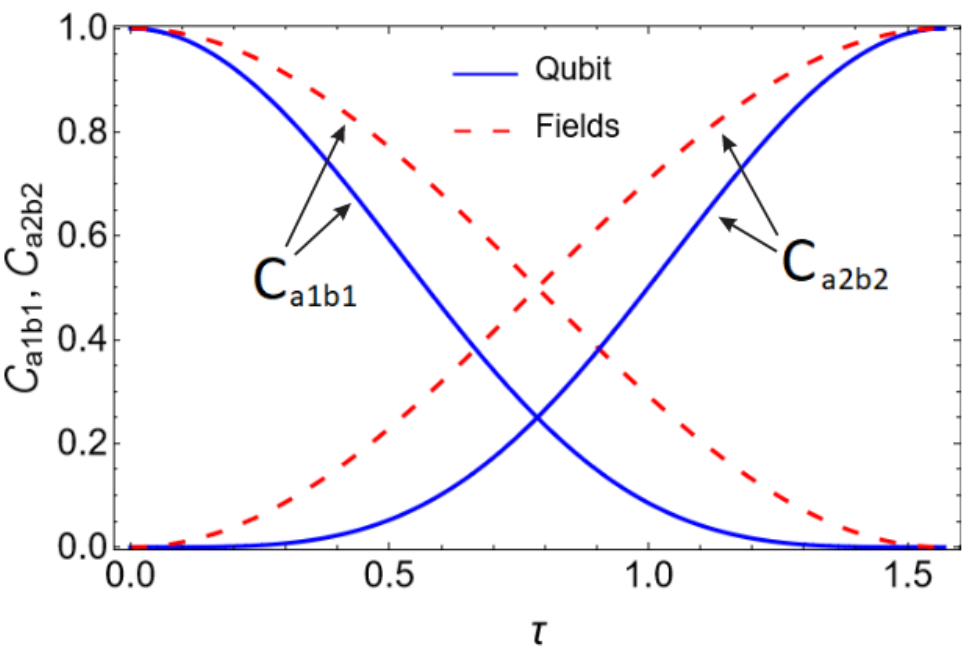}}\qquad\qquad
	\subfigure[\label{fig:2b}]{\includegraphics[scale=0.5]{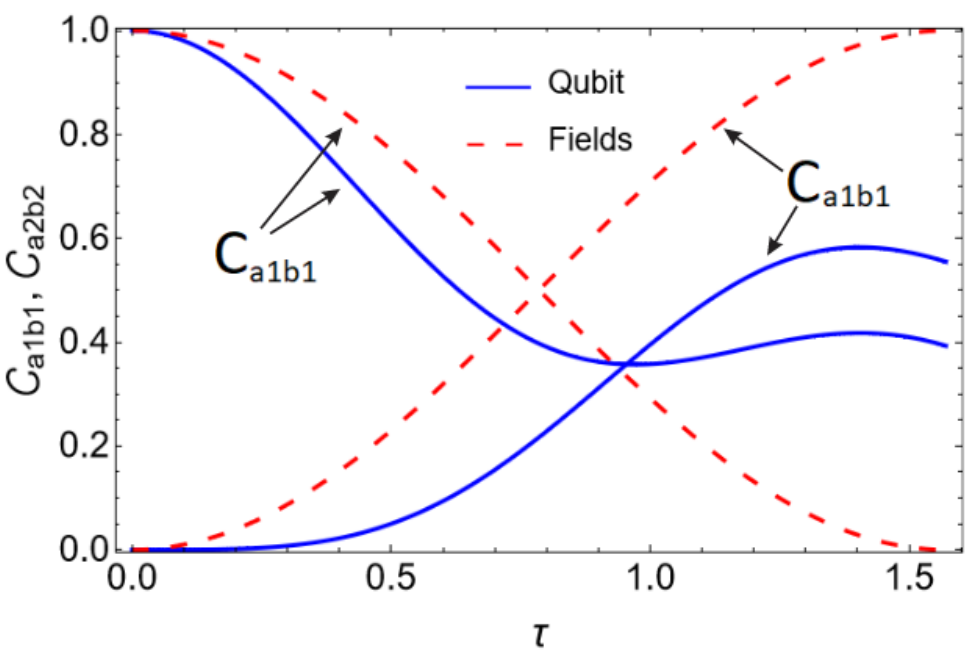}}\qquad
	\caption{Concurrence relative to the cavity $1$ ($a_1 \wedge b_1$) and cavity 2 ($a_2 \wedge b_2$) 
		modes, as a function of the normalized time $\tau$, for couplings via: a bridge qubit (blue line, $\tau^{(g)}$); 
		evanescent fields (dashed red line, $\tau^{(\lambda)}$). 
		Intra-cavity modes are uncoupled ($J_1=J_2= 0$) in (a), and coupled in (b), with $J_1=J_2=2\eta$. Here 
		$\eta=g$ refers to the bridge qubit coupling, and $\eta=\lambda$ to the coupling via evanescent waves.}
	\label{fig2}
\end{figure}

Now we are going to find at which times the initial entanglement of the intra-cavity modes in cavity $1$ 
(between the $a_1 \wedge b_1$ field modes) is fully transferred to the field modes in cavity $2$ ($a_2 \wedge b_2$). 
For that, we employ the Concurrence function \cite{wootters98}, a widely used quantifier of entanglement defined as

\begin{equation}\label{concurrence}
	C(\rho_j) = \textit{max} \left\lbrace 0, \sqrt{\lambda_1} - \sqrt{\lambda_2} - \sqrt{\lambda_3} - \sqrt{\lambda_4}  \right\rbrace,
\end{equation}
where $\lambda_i$ are the eigenvalues in decreasing order of the non-Hermitian matrix $\rho_j \tilde{\rho}_j$ 
of the corresponding bipartitions, with $\tilde{\rho}_j = (\sigma_y \otimes \sigma_y) {\rho^{*}_j} (\sigma_y \otimes \sigma_y)$. 
The index $j$ may refer to the density matrix of the system of two cavities coupled via: the qubit $(j = g)$; or evanescent waves 
$(j = \lambda)$. For instance, the Concurrence of the bipartite state constituted by the fields of cavity 1 is
obtained from the reduced density operator $\rho_j \equiv \rho^{(a1,b1)}_j$, after tracing over the remaining sub-systems, i.e.,
$\rho^{(a1,b1)}_g = \mbox{Tr}_{a2,b2,q}\Big[ |\Psi\rangle_g{}_g\langle\Psi |\Big]$ (bridge qubit coupling) and 
$\rho^{(a1,b1)}_\lambda = \mbox{Tr}_{a2,b2}\Big[ |\Phi\rangle_\lambda{}_\lambda\langle\Phi |\Big]$ (evanescent fields coupling).
For convenience, we calculated the Concurrence as a function of the normalized time $\tau$, with
correspondence to each type of coupling defined as $\tau^{(j)}$: i) $\tau^{(g)} \equiv gt$ (qubit); and 
ii) $\tau^{(\lambda)} \equiv \lambda t$ (evanescent fields).
We should stress that our calculations are compatible with typical experimental values of the relevant parameters \cite{oshea13,dayan1062}, 
that is: $g/2\pi = 70$ MHz, $\lambda/2\pi = 30$ MHz, and $J_{1,2}$ may vary from $J_{1,2} = 0$ to $J_{1,2}/2\pi = 250$ MHz.
In Fig. \ref{fig2} we have plots of the Concurrence as a function of the normalized time $\tau$, in relation to 
the field modes in cavity $1$ ($a_1 \wedge b_1$) as well as as the field modes in cavity $2$ ($a_2 \wedge b_2$). In Fig. \ref{fig:2a} we 
considered uncoupled intra-cavity modes, i.e., $J_1 = J_2 = 0$. In this case, we find that regardless of the type of coupling between cavities,
at the specific time $\tau_2 = \pi/2$ the field modes in cavity $1$ are non-entangled $(C_{a1b1} = 0)$, while the field modes in cavity 
$2$ become maximally entangled, $(C_{a2b2} = 1)$, i.e., entanglement is fully transferred from cavity $1$ to cavity $2$.
However the degree of entanglement for intermediate times is slightly different in each case, as shown in Fig. \ref{fig:2a}. 
If the intra-cavity field couplings are nonzero ($J_{1,2} \neq 0$), maximal entanglement in cavity $2$ is still achieved in 
the evanescent fields case, although this does not occur for the coupling via a bridge qubit, as seen in Fig. \ref{fig:2b}.

\begin{figure}[htpb]
	\centering
	\includegraphics[scale=0.4]{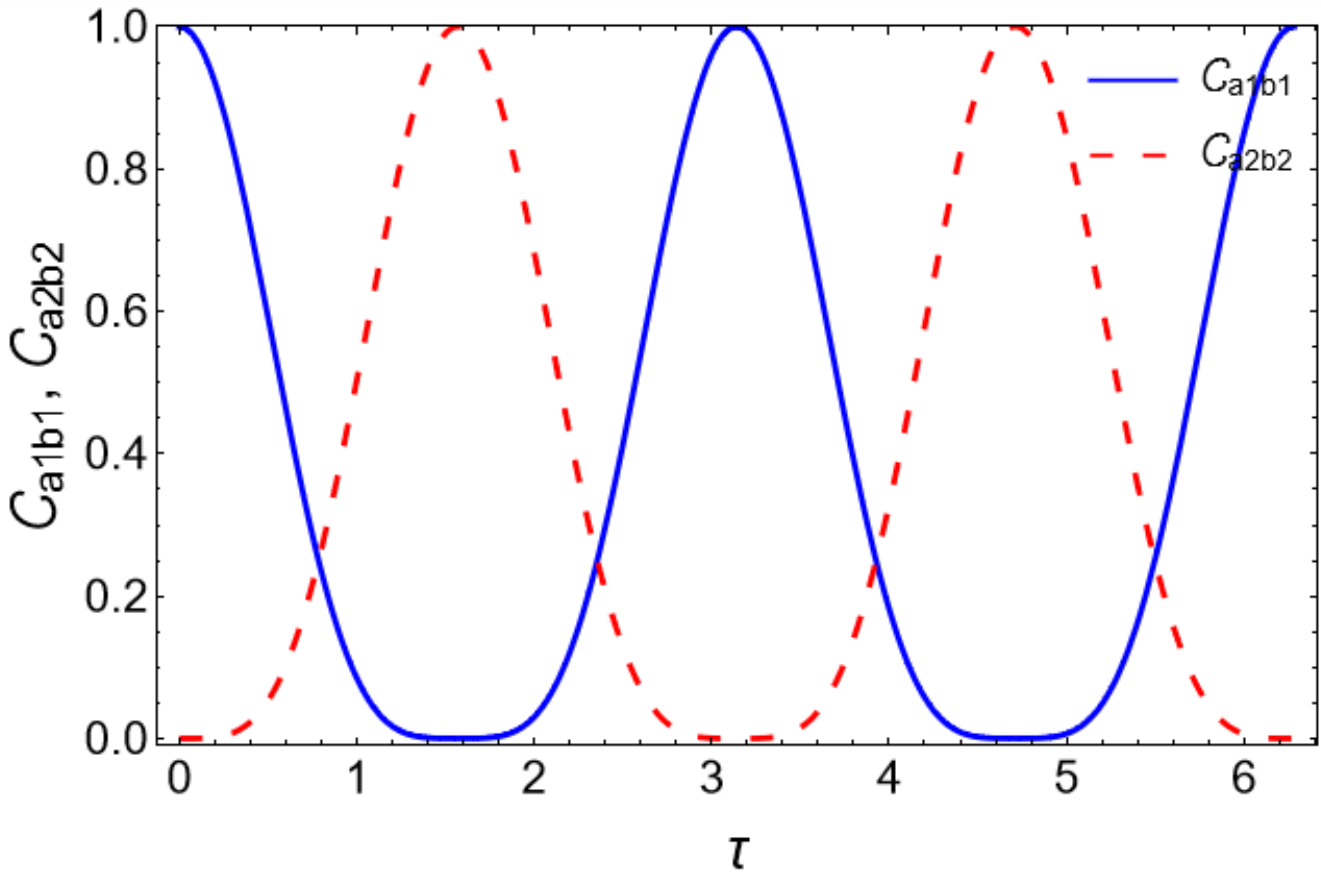}
	\caption{Concurrence as a function of the normalized time $\tau$ (showing ``sudden death") relatively to the field modes in cavity $1$ ($a_1\wedge b_1$) 
		and in cavity $2$ ($a_2\wedge b_2$) for a bridge qubit coupling, with $g_1=g_2=g$ and $J_1=J_2=0$. The initial state of the system is
		$\vert {\psi (0)}\rangle =\frac{1}{\sqrt{2}} (\vert 10\rangle + \vert 01\rangle)_{c1}\vert g \rangle \vert 00 \rangle_{c2}$.}
	\label{fig3}
\end{figure}

We would like now to take a more careful look at the bridge qubit coupling case for non-interacting intra-cavity modes ($J_{1,2} = 0$) 
having an initial maximally entangled state in cavity $1$. For that we plot the Concurrence as a function of the normalized time $\tau$, 
relatively to cavity $1$ modes as well as cavity $2$ modes for two full periods (up to $\tau = 2\pi$), as shown in Fig. \ref{fig3}. 
We clearly see that while entanglement is transferred from cavity $1$ to cavity $2$ at $\tau_2 = \pi/2$, the cavity $1$ modes are kept 
with null entanglement during a finite time interval, which characterizes the phenomenon known as Sudden Death of Entanglement (SDE) 
\cite{horodecki01,eberly04}.The entanglement goes down to zero before $\tau_2$ and revives after a short time, i.e., a Sudden Rebirth 
of Entanglement (SRE) occurs. We should point out that both SDE and SRE can take place even if the system is not coupled to an environment 
with many degrees of freedom, as discussed in \cite{decordi20}. Since the evolution of entanglement is periodic, the Concurrence 
of the field modes in cavity $1$ 
returns to its maximum value $C = 1.0$ at $\tau_3 = \pi$. However, the entanglement with respect to the field modes $a_2 \wedge b_2$ 
goes to zero before $\tau_3 = \pi$, remaining null for a finite time interval, i.e., SDE also takes place in the field modes of cavity $2$, 
as seen in Fig. \ref{fig3}. Yet, the SDE/SRE does not occur in a coupling via evanescent fields, 
emerging only if the cavities are connected via a bridge qubit. 

\subsection{Entangled state generation}

In this section we are going to discuss the generation of entangled states involving the sub-systems available: 
the qubit (in the case of coupling via a qubit), and the WGMs. 

\subsubsection{Coupling via a bridge qubit}

Firstly we discuss the generation of entangled states involving the propagating and counter-propagating modes 
of distinct cavities and the bridge qubit. In this case we assume an initial preparation of the system in the 
state $\vert \Psi(0)\rangle = (\cos \theta\vert 10\rangle + \sin \theta \vert 01\rangle)_{c1}\vert g\rangle \vert 00\rangle_{c2}$. Then, 
if the evolution of the system is governed by the Hamiltonian in Eq. (\ref{eq:Hq})  
from an initial product state $(\theta = 0)$ with non-interacting intra-cavity modes $(J_{1,2} = 0)$, we find 
that the state of the system at a time $\tau_2 = \pi/2$ will be the following 4-partite entangled state  

\begin{eqnarray}
	&&\vert \Psi (\tau=\pi/2, \theta=0, J_{1,2}=0)\rangle = \nonumber \\ 
	&&=\frac{1}{2}\big(\vert 1\rangle_{a1} \vert 0\rangle_{b1} \vert 0\rangle_{a2} \vert 0\rangle_{b2} + \vert 0\rangle_{a1} \vert 1\rangle_{b1} \vert 0\rangle_{a2} \vert 0\rangle_{b2} \nonumber \\ 
	&&+\vert 0\rangle_{a1} \vert 0\rangle_{b1} \vert 1\rangle_{a2} \vert 0\rangle_{b2} 
	+ \vert 0\rangle_{a1} \vert 0\rangle_{b1} \vert 0\rangle_{a2} \vert 1\rangle_{b2})\vert g\rangle.
	\label{eq:qubit:J=0}
\end{eqnarray}

In other words, the generated qubit-assisted 4-partite state in Eq. (\ref{eq:qubit:J=0}) involves the four field modes, 
$a_1$, $b_1$, $a_2$ and $b_2$, leaving the state of the qubit factored out. The generation of a 4-partite field
state in the two cavity system is made possible because the bridge qubit couples equally to each mode of cavity $1$ as well as 
to each mode of cavity $2$, as seen in Eq. (\ref{hamilqubitcoup}). Thus, we expect that there will be a time, during evolution, 
when all four modes of the field will become entangled, thanks to the symmetry of the coupling between the two cavities.

\begin{figure}[htpb]
	\centering
	\subfigure[\label{fig:4a}]{\includegraphics[scale=0.4]{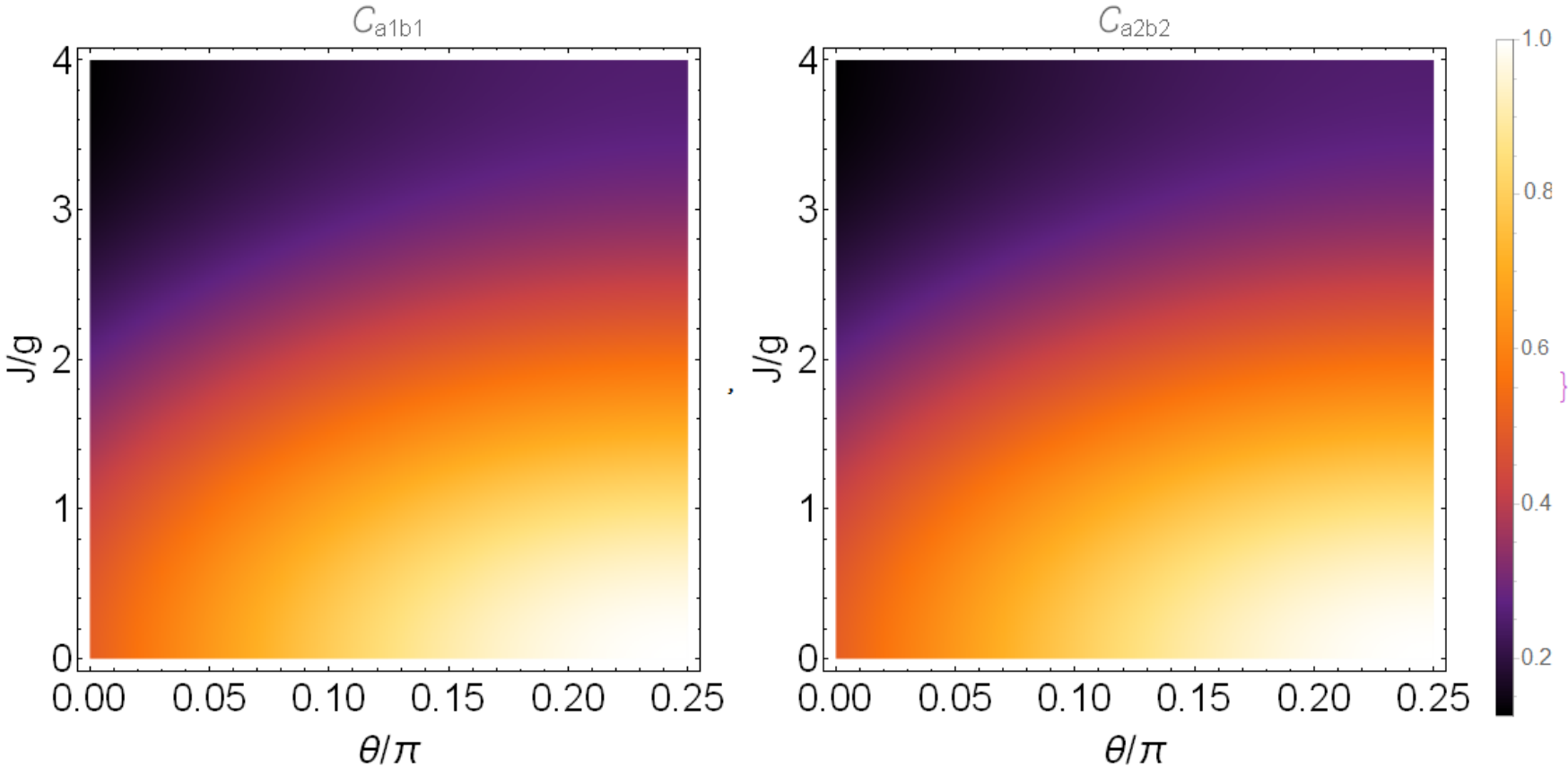}}\qquad
	\subfigure[\label{fig:4b}]{\includegraphics[scale=0.4]{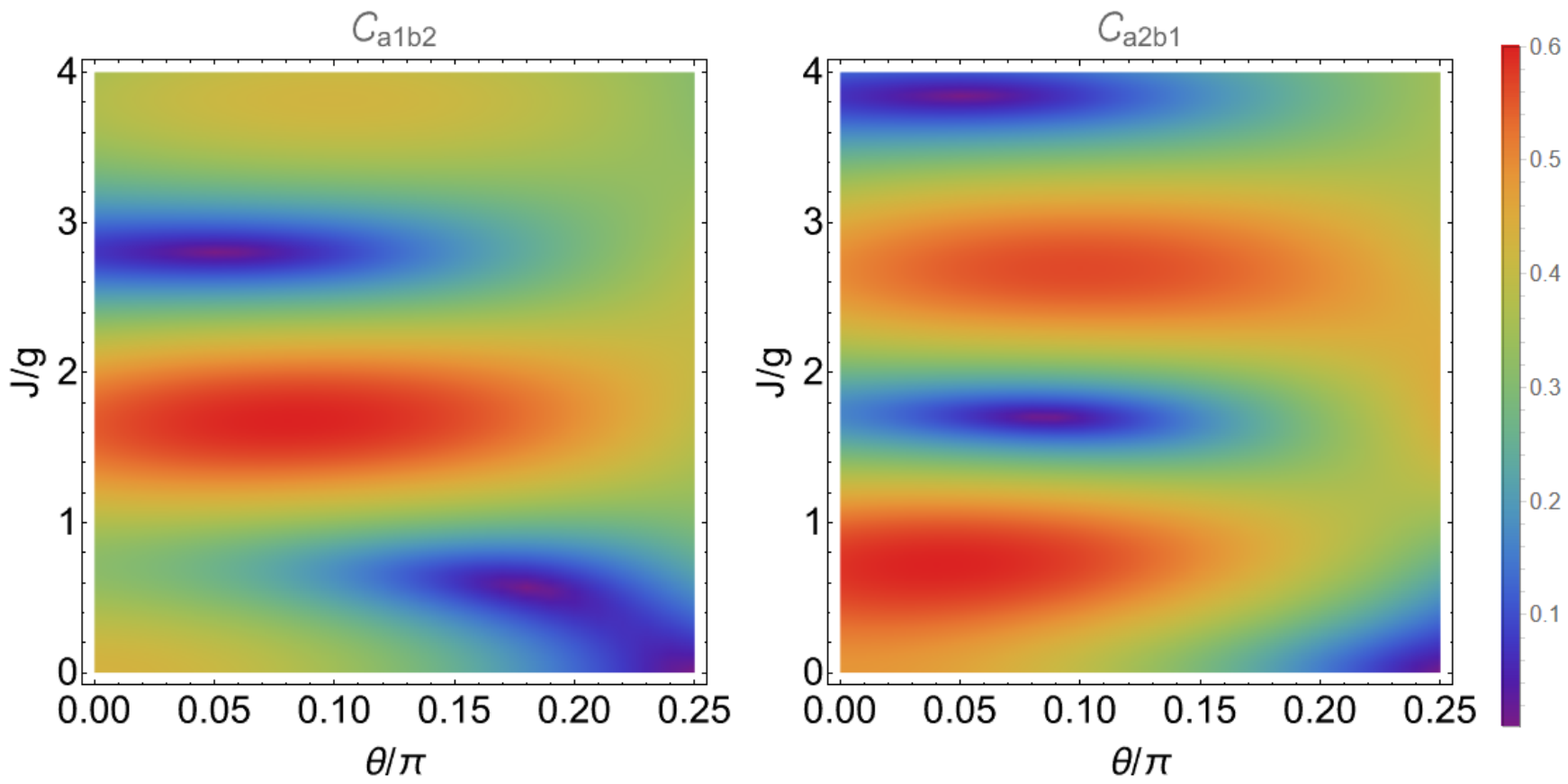}}\qquad
	\caption{Bridge qubit coupling. The plots above show the Concurrence for specific 
		pairs of field modes, at time $\tau_2=\pi/2$, as a function of $J\equiv J_{1,2}$ and 
		the degree of entanglement of the initial states in cavity $1$ ($\theta$). a) Concurrence relative to the 
		intra-cavity modes, $a_1 \wedge b_1$ and $a_2 \wedge b_2$; b) Concurrence relative to the inter-cavity modes, 
		$a_1 \wedge b_2$, $b_1 \wedge a_2$. The entanglement between modes $a_1 \wedge a_2$ and 
		$b_1 \wedge b_2$, not shown here, features an identical pattern.} 
	\label{fig4}
\end{figure}

We could now analyse the bipartite entanglement between various subsystems as a function of both the intra-cavity modes 
interaction $(J_{1,2})$ and the degree of entanglement of the initial state associated to cavity $1$ (parameter $\theta$). 
In Fig. \ref{fig4} we have plots of the Concurrence at the specific time $\tau_2 =\pi/2$, as a function of $J\equiv J_1 = J_2$ 
and $\theta$. Firstly we note that if the cavities are coupled via a bridge qubit, it is possible (at $\tau_2= \pi/2$) to have 
maximally entangled states ($C = 1$) involving the intra-cavity modes ($a_1 \wedge b_1$ and $a_2 \wedge b_2$), 
only if the fields are initially in an entangled state $(\theta \neq 0)$, as seen in Fig. \ref{fig:4a}. 
On the other hand, the modes $a_1 \wedge b_2$ and $a_2 \wedge b_1$ (and also $a_1 \wedge a_2$ and $b_1 \wedge b_2$) never 
become maximally entangled, irrespective of the initial state, reaching at most a Concurrence of $C \approx 0.5$, as shown 
in Fig. \ref{fig:4b}.

\begin{figure}[htpb]
	\centering
	\includegraphics[scale=0.4]{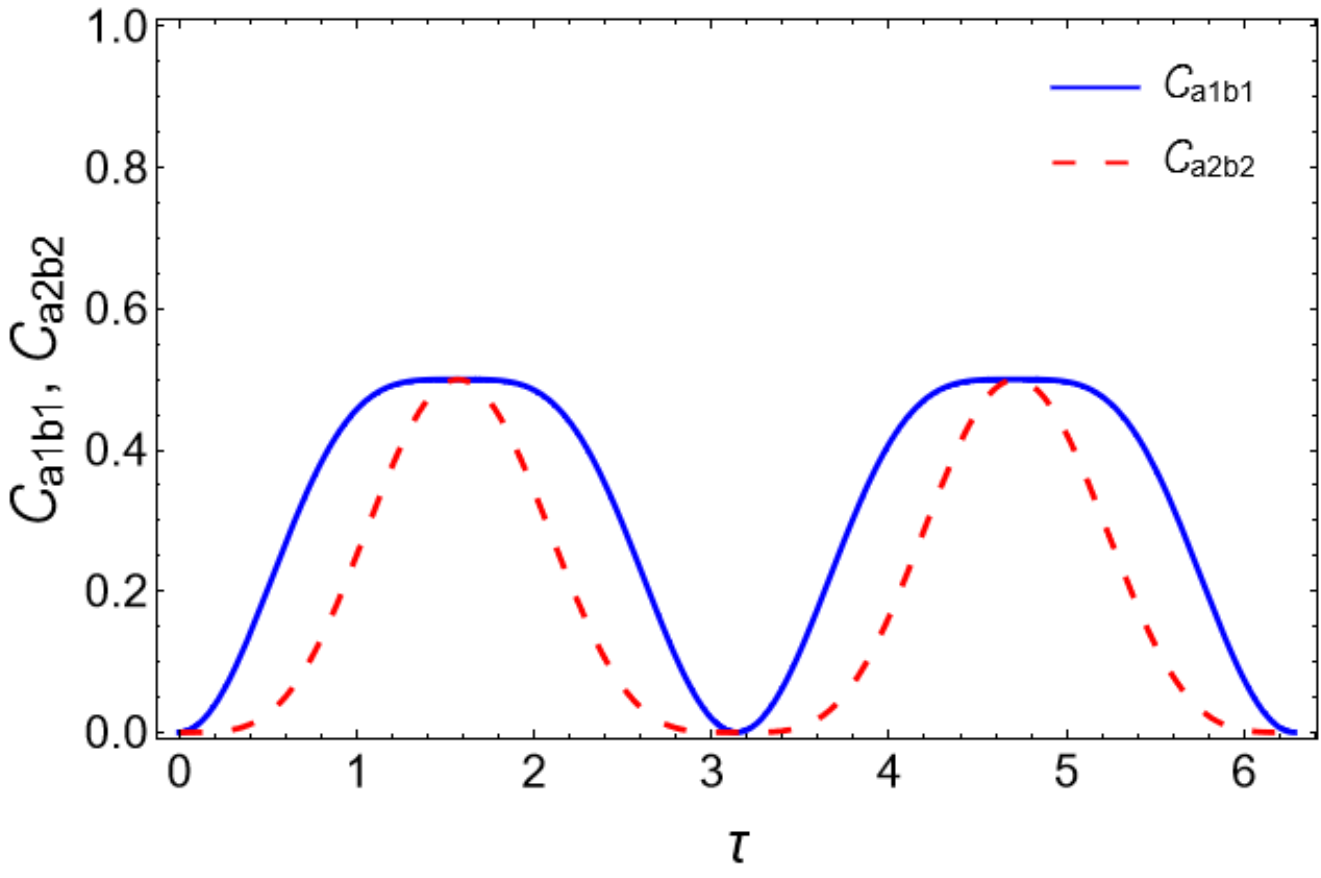}
	\caption{Concurrence as a function of the normalized time $\tau$ (showing ``sudden freezing'') relatively to the field modes 
		in cavity $1$ ($a_1 \wedge b_1$) and in cavity $2$ ($a_2 \wedge b_2$) for a bridge qubit coupling, with $g_1=g_2=g$ and $J_1=J_2=0$. 
		The initial state of the system is $\vert {\psi (0)}\rangle =\frac{1}{\sqrt{2}} \vert 10\rangle_{c1}\vert g \rangle \vert 00 \rangle_{c2}$.}
	\label{fig5}
\end{figure}

We can also look closer at the evolution of entanglement considering an initial preparation in the separable state 
$\vert \psi(0)\rangle = \vert 10\rangle_{c1}\vert g\rangle\vert 00\rangle_{c2}$. Fig. \ref{fig5} illustrates the evolution of entanglement 
between the intra-cavity modes of both cavities up to the time $\tau = 2\pi$. Note that, unlike the case of 
having an initial entangled state prepared in cavity $1$, here the entanglement between modes $a_1 \wedge b_1$ (and between
modes $a_2 \wedge b_2$) reaches the maximum value of $C = 0.5$ at $\tau_2 = \pi/2$. This happens because precisely at that specific 
time, all four fields constitute a 4-partite entangled state, leaving the atom separated. As a consequence, the bipartite states 
formed by the fields in cavity $1$ $(a_1 \wedge b_1)$ and in cavity $2$ $(a_2 \wedge b_2)$ are each one in a partially entangled 
state at $\tau_2 = \pi/2$, with $C = 0.5$. We also note that the entanglement evolution in 
each cavity shows different patterns; in cavity $1$, the maximum entanglement is reached more quickly than in cavity $2$, 
and it also decays more slowly. We also observe that the entanglement in cavity $1$ is kept at a constant (maximum) value 
during a significantly large time interval. This resembles the so-called ``freezing (and thawing) of entanglement" \cite{rau14,eberly21}, a 
phenomenon in which entanglement ``freezes" at some particular value during the evolution, decreasing after some time. 
On the other hand, in cavity $2$ the entanglement between modes $a_2 \wedge b_2$ undergoes
SDE, becoming null in a finite time interval around $\tau_3 = \pi$. Nonetheless, the field modes in cavity $1$ do not suffer SDE.
Clearly, the entanglement distribution in the system as well as its ``freezing" depend on which subsystem contains the initial field state excitation
\cite{rau14,eberly21}. If the photon initially populates mode $a_1$, entanglement freezing occurs in cavity $1$. Conversely,
the modes in cavity $2$ undergo SDE. However, if the photon populates mode $a_2$ instead, it will be the other way around, i.e., 
freezing will occur in in cavity $2$, while SDE will take place in cavity $a_1$. 

\subsubsection{Coupling via evanescent waves}

In the case of having the cavities connected via evanescent waves, a direct coupling between field 
modes is established, and we may assume an initial state of the form 
$\vert \Phi(0)\rangle =(\cos \theta\vert 10\rangle + \sin \theta \vert 01\rangle)_{c1}\vert 00\rangle_{c2}$, involving only field states. 
Having the evolution of the system governed by the Hamiltonian in Eq. (\ref{eq:Hlambda}) from an initial product state 
$(\theta = 0)$ and non-interacting intra-cavity modes $(J_i = 0)$, we find that at $\tau_1 = \pi/4$ the following entangled 
state is generated,
\begin{equation}
	\vert \Phi (\tau=\pi/4, \theta=0, J_{1,2}=0)\rangle = \frac{1}{\sqrt{2}}(\vert 1\rangle_{a1}\vert 0\rangle_{b2} + \vert 0\rangle_{a1}\vert 1\rangle_{b2}) 
	\otimes \vert 0\rangle_{b1}\vert 0\rangle_{a2},
	\label{eq:campos:J=0}
\end{equation}
i.e., a maximally entangled state with modes $a_1 \wedge b_2$. At the same time, the modes $a_2 \wedge b_1$ remain non-entangled. 
Interestingly, if we switch on the interaction between the intra-cavity modes in the cavities $(J_1 = J_2 = 2\lambda)$, a different 
entangled state also involving the modes in both cavities will be formed,
\begin{equation}
	\vert \Phi (\tau=\pi/4, \theta=0, J_{1,2}=2\lambda)\rangle = \frac{1}{\sqrt{2}}(\vert 1\rangle_{b1}\vert 0\rangle_{a2} 
	+ \vert 0\rangle_{b1}\vert 1\rangle_{a2}) \otimes \vert 0\rangle_{a1}\vert 0\rangle_{b2},
	\label{eq:campos:J=2}
\end{equation}
which is a maximally entangled state of modes $a_2 \wedge b_1$, remaining the modes $a_1 \wedge b_2$ non-entangled. 

\begin{figure}[htpb]
	\centering
	\subfigure[\label{fig:6e}]{\includegraphics[scale=0.4]{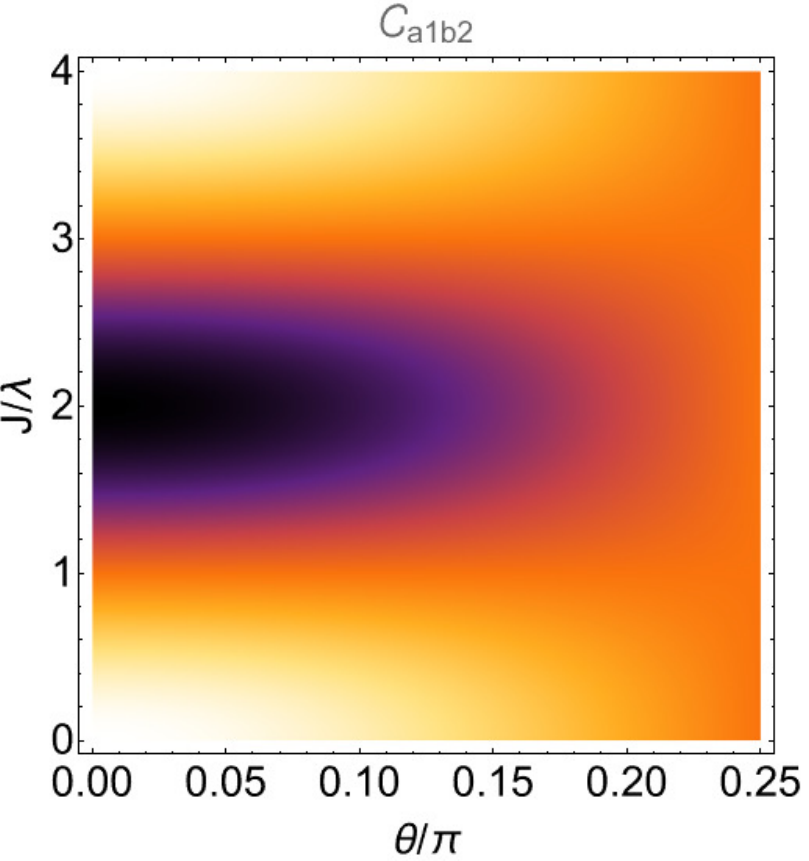}}
	\subfigure[\label{fig:6f}]{\includegraphics[scale=0.4]{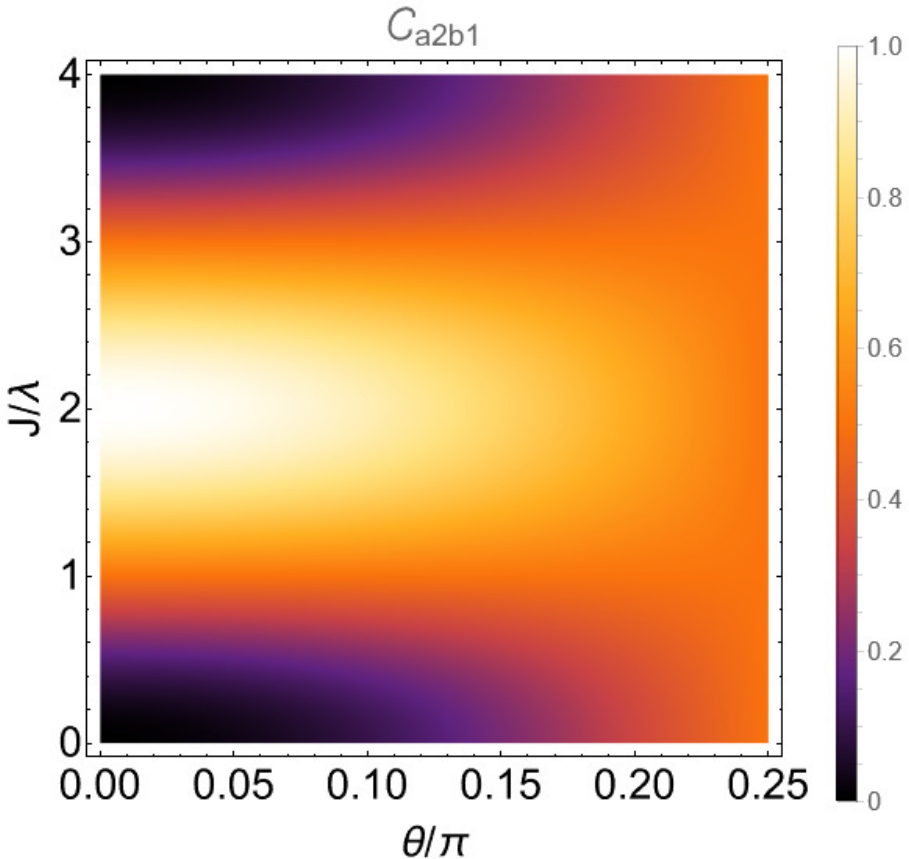}}\qquad
	\caption{Evanescent waves coupling. The plots above show the Concurrence for specific 
		pairs of field modes, at $\tau_1=\pi/4$, as a function of the $J\equiv J_{1,2}$ coupling and the degree of entanglement of 
		the initial states in cavity $1$ ($\theta$);  a) modes $a_1 \wedge b_2$; and b) modes $a_2 \wedge b_1$.}
	\label{fig:campos_todos}
\end{figure}

Hence, the generation of specific entangled states between different modes (Eq. (\ref{eq:campos:J=0}) 
and Eq. (\ref{eq:campos:J=2})) 
can be accomplished simply by tuning the intra-cavity couplings $J_i$. 
We may also calculate the Concurrence by fixing the time, $\tau_1 = \pi/4$, and varying the couplings $J_i$ as well as the 
weighting angle $\theta$, as shown in Fig. \ref{fig:campos_todos}. It is clear that the Concurrence is a periodic function of 
the parameters $J_i$, exhibiting different patterns which will depend on the pair of field modes considered. We remark that 
maximum entanglement $(C = 1)$ between modes of distinct cavities is achieved (at $\tau_1 = \pi/4$) for specific values 
of the couplings $J_i$, namely, for $J_{1,2} = 0$ and $J_{1,2} = 4\lambda$ in the case of $a_1 \wedge b_2$, 
and $J_{1,2} = 2\lambda$ in the case of $a_2 \wedge b_1$. Moreover, the 
generation of entanglement is possible from an initial joint product state, and that an initial maximally
entangled state in cavity $1$ $(\theta = \pi/4\, \mbox{rad})$ does not result in a maximally state of other modes at $\tau_1 = \pi/4$. Yet, for each of the remaining bi-partitions ($a_1 \wedge b_1$, $a_2 \wedge b_2$, $a_1 \wedge a_2$, 
and $b_1 \wedge b_2$), the maximum possible value of the Concurrence is $C = 0.5$, irrespective of the values of $J_i$ 
and $\theta$.

\newpage
\subsection{Coupling via an optical fiber}

Another possible way of coupling the microtoroidal cavities is via an optical fiber. We assume
the so-called short fiber limit, in which the cavities couple effectively only to a single mode of the fiber, here
described by creation and annihilation operators $c^\dagger$ and $c$, respectively. The complete Hamiltonian including the 
fiber reads:
\begin{eqnarray}
	H_{T} = H_{S} + H_{\nu},\label{eq:Hf}
\end{eqnarray}
where
\begin{equation}
	H_{\nu} = \hbar \nu \left[( a_{1}^{\dagger} + b_{1}^{\dagger}) c + (a_{1} + b_{1}) c^\dagger 
	+ ( a_{2}^{\dagger} + b_{2}^{\dagger}) c + (a_{2} + b_{2}) c^\dagger\right] 
	\label{hamilfibercoup}
\end{equation}
is the effective interaction term of the cavity modes ($a_i,b_i$) with the fiber mode ($c$) with coupling 
constant $\nu$. 
Due to the analogy between the Hamiltonians in Eqs. (\ref{hamilqubitcoup}) and  (\ref{hamilfibercoup}),
equivalent results will be obtained in cavity systems with couplings either via a fiber or via a qubit, 
provided that the field modes involve only one excitation (one photon).
\vspace{1cm}

\section{\label{sec:dissip} Effects of losses}

In order to make a more realistic description of the dynamics of the cavity system, we will now discuss the influence 
of cavity losses and atomic spontaneous emission on the bipartite entanglement of the intra-cavity field modes. 
Both effects can be modelled via the usual GKSL-type master equation for the reduced 
density operator $\rho_j \equiv \rho^{(al,bl)}_j$ describing the bipartite state relative to the $l$-th cavity
\begin{eqnarray}\label{mastereq}
	\frac{d \rho_j}{dt} &=& -\frac{\textit{i}}{\hbar} [H_{j},\rho_j] + \kappa \sum_{i=1,2}\left(2 a_{i} 
	\rho_j a_{i}^\dagger - a_{i}^\dagger a_i \rho_j - \rho_j a_{i}^\dagger a_i\right) \\  
	&+& \kappa\sum_{i=1,2}\left(2 b_{i} \rho_j b_{i}^\dagger - b_{i}^\dagger b_i \rho_j - \rho_j b_{i}^\dagger b_i\right) + 
	\gamma\Big(2 \sigma_- \rho_j \sigma_+ - \sigma_+ \sigma_- \rho_j - \rho_j \sigma_+ \sigma_-\Big). \nonumber
\end{eqnarray}
The index $j$ is related to couplings with the qubit $(j = g)$ or via evanescent waves $(j = \lambda)$, respectively. The first
term in the right-hand side of Eq. (\ref{mastereq}) refers to the unitary part, with evolution governed by the Hamiltonian
$H_{j}$. The other three terms (Lindbladians) represent the action of external zero temperature reservoirs ($T = 0$K) which in 
general degrade the quantum state properties. For simplicity we assume the same dissipation rate $\kappa$ for each field mode 
in each of the cavities. In the case of coupling via a qubit, we also take into account the spontaneous emission at a rate 
$\gamma$, which is typically smaller than $\kappa$ \cite{dayan1062}. Of course
the last term in Eq. (\ref{mastereq}) does not exist if the cavities are coupled through evanescent fields.
By numerically solving Eq. (\ref{mastereq}) for the reduced density operator $\rho_j$, one may calculate
the Concurrence for various bipartite states of cavity fields using Eq. (\ref{concurrence}). The results
will be presented below.

\begin{figure}[htpb]
	\centering
	\subfigure[\label{fig:2ax}]{\includegraphics[scale=0.4]{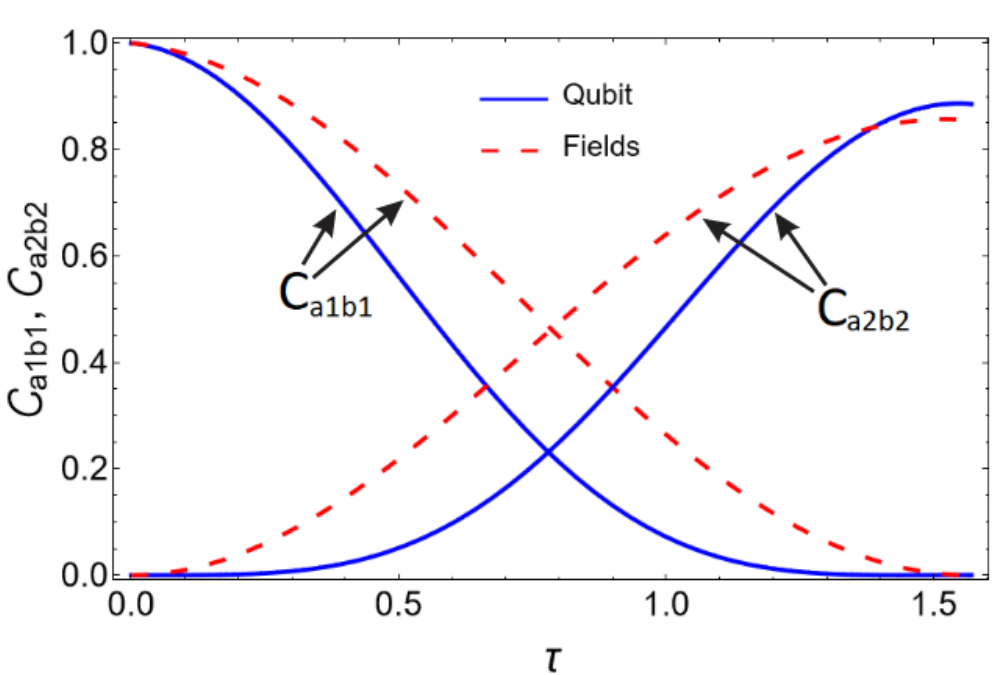}}\qquad
	\subfigure[\label{fig:2bx}]{\includegraphics[scale=0.4]{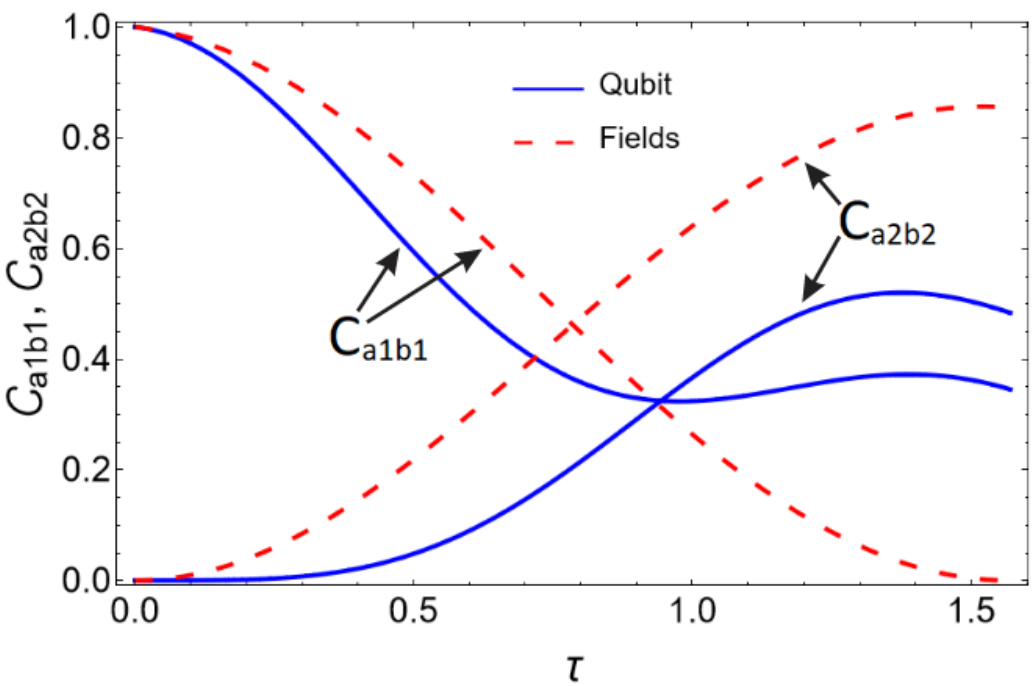}}\qquad
	\caption{Effects of losses on quantum entanglement. Concurrence as a function of the normalized time $\tau$ of the cavity 
		$1$ ($a_1 \wedge b_1$) and cavity 2 ($a_2 \wedge b_2$) modes, for couplings via: a bridge qubit (blue line, $\tau^{(g)}$); evanescent fields 
		(dashed red line, $\tau^{(\lambda)}$). The field modes and the qubit are damped at a rate $\kappa = 5 \times 10^{-2}\eta$ and $\gamma = 5 \times 10^{-3}\eta$, respectively. For uncoupled intra-cavity modes ($J_1=J_2= 0$) in (a), and coupled modes in (b): $J_1=J_2=2\eta$. Here $\eta=g$ refers to the bridge qubit coupling, and $\eta=\lambda$ to the coupling via evanescent waves.}
	\label{fig12}
\end{figure}

\begin{figure}[htpb]
	\centering
	\includegraphics[scale=0.4]{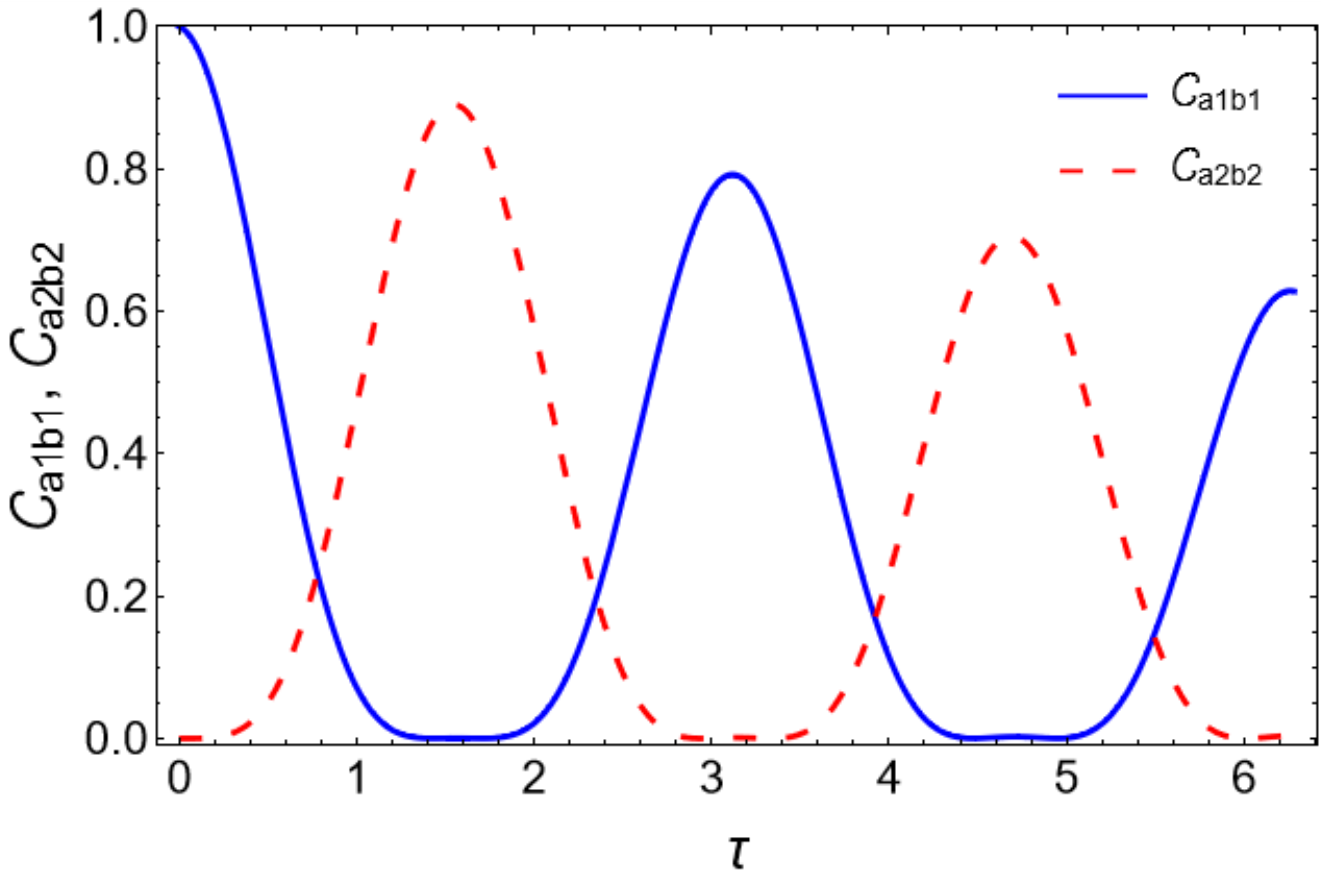}
	\caption{Effects of losses on quantum entanglement. Concurrence as a function of the normalized time $\tau$ relatively to the field modes in cavity $1$ ($a_1\wedge b_1$) 
		and in cavity $2$ ($a_2\wedge b_2$) for a bridge qubit coupling, with $g_1=g_2=g$ and $J_1=J_2=0$. The field modes and the qubit are damped at a rate $\kappa = 5 \times 10^{-2}g$ and $\gamma = 5 \times 10^{-3}g$, respectively. The initial state of the system is $\vert {\psi (0)}\rangle =\frac{1}{\sqrt{2}} 
		(\vert 10\rangle + \vert 01\rangle)_{c1}\vert g \rangle \vert 00 \rangle_{c2}$.}
	\label{fig13}
\end{figure}

\begin{figure}[htpb]
	\centering
	\includegraphics[scale=0.4]{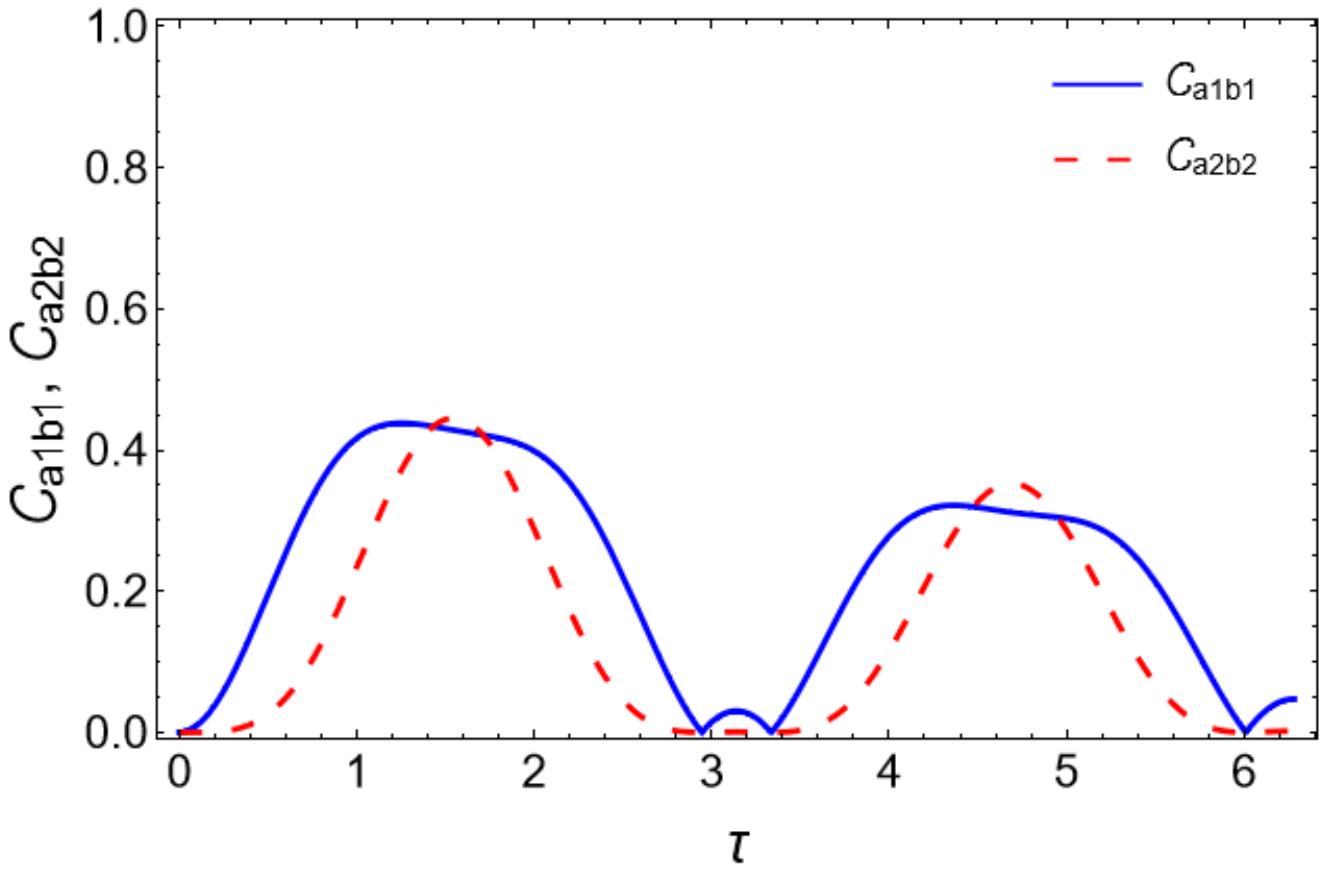}
	\caption{Effects of losses on quantum entanglement. Concurrence as a function of the normalized time $\tau$ relatively to the field modes in cavity $1$ 
		($a_1 \wedge b_1$) and in cavity $2$ ($a_2 \wedge b_2$) for a bridge qubit coupling, with $g_1=g_2=g$ and $J_1=J_2=0$. 
		The field modes and the qubit are damped at a rate $\kappa = 5 \times 10^{-2}g$ and $\gamma = 5 \times 10^{-3}g$, respectively. 
		The initial state of the system is $\vert {\psi (0)}\rangle =\frac{1}{\sqrt{2}} \vert 10\rangle_{c1}\vert g \rangle \vert 00 \rangle_{c2}$.}
	\label{fig14}
\end{figure}

Regarding entanglement transfer, we found that losses have slightly different effects depending on the type of coupling. 
In Fig. \ref{fig12}(a) we note that although the 
complete transfer of entanglement from one cavity to another is precluded by losses, the maximum attained entanglement 
at $\tau_2 \approx \pi/2$ is larger in the case of coupling via a bridge qubit, compared to the coupling via 
evanescent waves. This is because the qubit's spontaneous emission rate is much smaller than the cavity decay rate, and therefore 
it is less likely a photon to be lost if the cavities are coupled through a qubit rather than via evanescent waves. 
We recall that for non-zero intra-cavity couplings ($J_1=J_2 \neq 0$), even under ideal conditions 
the full transfer of entanglement between cavities is not possible if they are coupled by a qubit. The losses of 
course makes the situation even worse, as shown in Fig. \ref{fig12}(b). In Fig. \ref{fig13} we have plotted the evolution of 
Concurrence up to $\tau = 2\pi$ to show that, in addition to the expected drop in the maximum values of entanglement, we observe 
that the time interval in which the entanglement is zero (sudden death), gets a little longer as time passes. The freezing 
of entanglement will also suffer the effects of losses, as shown in Fig. \ref{fig14}. Interestingly, we notice 
small entanglement ``revivals" around $\tau_3 = \pi$ (and multiples), that is, at times when the Concurrence is null 
in the absence of losses.

\newpage
\section{\label{sec:conclu} Conclusions}

In summary, we have performed a theoretical study demonstrating the experimental feasibility of entanglement 
generation as well as transfer of maximally entangled states in a system comprising two identical WGMs microtoroidal resonators. 
We discuss two different types of inter-cavity couplings, that is, using a bridge qubit or via evanescent waves (directly).
Firstly we have shown that maximally entangled states of two modes belonging to the same cavity can be transferred with 
$100\%$ fidelity to the other cavity (in the ideal case). Regarding the generation of entangled states from initial product states, 
each one of the couplings leads to different entangled states. For instance, we have found that for a
evanescent waves coupling, it is allowed the generation of maximally bipartite entangled states (Concurrence $C = 1.0$) for modes 
belonging to different cavities, which manifests the non-local behavior of the resulting quantum states.  
Besides, we have found that even if the microtoroidal resonators are structurally deformed (leading to $J_{1,2}\neq 0$), it is possible 
to generate maximally entangled states of pairs of field modes in cavities coupled via evanescent waves. 
Interestingly, for non-zero intra-cavity couplings, different pairs of field modes become entangled, in comparison to the states generated 
if $J_{1.2} = 0$. Another important result is related to the bridge qubit case, as it allows the generation of 
a 4-partite state of the cavity fields. Yet despite the fact that the processes of transfer and generation of 
entanglement are hindered by losses, in the bridge qubit coupling case the field entanglement is more robust against losses when compared 
to the direct coupling. Thus, the bridge qubit may act both as an assistant of entanglement distribution, as well as an 
obstacle to entanglement degradation in the two-cavity system. The present study is, therefore, valuable for understanding the several possibilities
of transfer and generation of entangled states in two WGMs cavities, depending on how they are coupled, which is certainly relevant for 
applications in quantum information processing tasks. In particular, the coupling via a bridge qubit turned out to be a very efficient way to 
distribute entanglement in that system, as well as to generate a unique 4-partite state of all four cavity fields.

\section*{Acknowledgments}

The authors would like to thank CNPq (Conselho Nacional de 
Desenvolvimento Cient\'\i fico e Tecnol\'ogico), Brazil,
for financial support through the National Institute for Science and 
Technology of Quantum Information (INCT-IQ under grant 465469/2014-0).

\bibliographystyle{elsarticle-num}
\bibliography{referenciass}

\end{document}